%% file: main.tex
\documentclass{article}
\usepackage[latin1]{inputenc}
\usepackage[T1]{fontenc}
\usepackage[english]{babel}
\usepackage{amssymb}
\usepackage{amsmath}
\usepackage{amsthm}
\usepackage{graphicx}
\usepackage[margin=3cm]{geometry}
\usepackage{hyperref}
\usepackage{comment}
\hypersetup{
	colorlinks=true,
	linkcolor=blue,
	filecolor=magenta,      
	urlcolor=cyan,
}
\usepackage[font=small,labelfont=bf]{caption}
\usepackage{multirow}
\usepackage{authblk}
\usepackage{gensymb}
\usepackage{booktabs}
\usepackage{cleveref}
\usepackage{subcaption}

\usepackage[
backend=biber,
style=numeric,
sorting=none
]{biblatex}

\title{Recovering the second moment of the strain distribution from neutron Bragg edge data}

\author[1,2]{Kyle Fogarty}
\author[3,1]{Evelina Ametova}
\author[4,1]{Genoveva Burca}
\author[5]{Alexander M. Korsunsky}
\author[6]{S\o{}ren Schmidt}
\author[7]{Philip J. Withers}
\author[1,*]{William R. B. Lionheart}

\affil[1]{Department of Mathematics, The University of Manchester, Manchester M13 9PL, United Kingdom}
\affil[2]{Department of Physics and Astronomy, The University of Manchester, Manchester M13 9PL, United Kingdom}
\affil[3]{Laboratory for Application of Synchrotron Radiation, Karlsruhe Institute of Technology, Karlsruhe 76131, Germany }
\affil[4]{ISIS Pulsed Neutron and Muon Source, STFC, UKRI, Rutherford Appleton Laboratory, Didcot OX11 0QX, United Kingdom}
\affil[5]{Department of Engineering Science, University of Oxford, Oxford OX1 3PJ, United Kingdom}
\affil[6]{European Spallation Source, Lund S - 221 00, Sweden}
\affil[7]{Henry Royce Institute, Department of Materials, The University of Manchester, M13 9PL, United Kingdom}
\affil[*]{Correspondence should be addressed to: \href{mailto:bill.lionheart@manchester.ac.uk}{bill.lionheart@manchester.ac.uk}}

\begin{document}
\maketitle

\vspace{5mm}
\begin{center} 
\normalsize{The following article has been submitted to Applied Physics Letters. \\After it is published, it will be found at \href{https://publishing.aip.org/resources/librarians/products/journals/}{Link}}
\end{center} 
\vspace{5mm}

\begin{abstract}
Point by point strain scanning is often used to map the residual stress (strain) in engineering materials and components. However, the gauge volume and hence spatial resolution is limited by the beam defining apertures and can be anisotropic for very low and high diffraction (scattering) angles. Alternatively, wavelength resolved neutron transmission imaging has a potential to retrieve information tomographically about residual strain induced within materials through measurement in transmission of Bragg edges ? crystallographic fingerprints whose locations and shapes depend on microstructure and strain distribution. In such a case the spatial resolution is determined by the geometrical blurring of the measurement setup and the detector point spread function. Mathematically, reconstruction of strain tensor field is described by the longitudinal ray transform; this transform has a non-trivial null-space, making direct inversion impossible. A combination of the longitudinal ray transform with physical constraints was used to reconstruct strain tensor fields in convex objects. To relax physical constraints and generalise reconstruction, a recently introduced concept of histogram tomography can be employed. Histogram tomography relies on our ability to resolve the distribution of strain in the beam direction, as we discuss in the paper. More specifically, Bragg edge strain tomography requires extraction of the second moment (variance about zero) of the strain distribution which has not yet been demonstrated in practice. In this paper we verify experimentally that the second moment can be reliably measured for a previously well characterised aluminium ring and plug sample. We compare experimental measurements against numerical calculation and further support our conclusions by rigorous uncertainty quantification of the estimated mean and variance of the strain distribution. 
\end{abstract}

\newpage
Residual stress (and thereby elastic strain) is the stress that remains in a body when no external forces are applied~\cite{withers2001residual}. Because these internal stresses add to those arising from externally applied loads, if they are not detected they can give rise to unexpected behaviours and premature failure. Therefore, information about the strain measured within polycrystalline materials is critically important for understanding the deformation and fracture mechanics of engineered components. A well-established technique used for strain measurements is based on neutron diffraction (or Bragg scattering). Depending on the material, the scattered neutrons will constructively interfere with each other only in particular directions and produce a intensity pattern (so-called Bragg peaks) from which the structure of the material is derived. Measurement of the position of Bragg peaks from diffraction allows the determination of lattice spacings, while the measurement of the relative shift in the positions provides information on lattice strains~\cite{fitzpatrick2003analysis}. To achieve high spatial resolution a sample is raster scanned with a collimated or focused beam and the angle of scattered beam $2\theta$ (\emph{e.g.} angle-dispersive diffraction) or wavelength/ energy (\emph{e.g.} energy-dispersive diffraction) is recorded to deduce the interplanar spacing, point by point, using Bragg's equation. This is then used to infer strain based on a comparison with the reference interplanar spacing. To overcome some of the disadvantages given by neutron diffraction measurements (\emph{e.g.} slow acquisition, the uncertainty of the exact specimen or gauge location along the beam~\cite{abbey2012neutron}) a new technique called Bragg edge neutron transmission for strain measurements was proposed and demonstrated\cite{Vogel_2000,steuwer2001sin2}. 

In this respect, a polychromatic neutron beam in a combination with a Time-of-Flight (ToF) area detector can be employed to register both spatial and ToF (wavelength) information about the transmitted neutrons. According to Bragg's law
\begin{equation}\label{eq:Bragg} 2 d_{hkl} \sin \theta =  \lambda_{hkl},
\end{equation} 
coherent elastic scattering at an incident angle of $\theta$ can happen only for wavelengths $\lambda$ shorter than twice the spacing between the lattice planes ($d_{hkl}$). Hence, the transmitted spectrum will exhibit a rapid increase in the transmitted intensity at a wavelength $\lambda$ slightly longer than twice this distance because intensity can no longer be diffracted out of the transmitted beam by this $hkl$ family of planes. This sharp change in transmission is called a \emph{Bragg edge} and allows the establishment of a relationship between the transmitted neutron spectral fingerprint and the crystallographic phases in the material. The application of the Bragg edge neutron transmission for strain mapping has been recently extended to high spatial resolutions due to advances in micro-channel-plate (MCP) detector technology~\cite{tremsin2012high,losko2021new}. 

Given a sample rotation, a strain tensor field in the object can be reconstructed tomographically (in general, rotations about six directions that do not lie on a projective conic are required to reconstruct tensor field~\cite{lionheart2015diffraction}). This technique is referred to as \emph{Bragg edge strain tomography} and seeks to determine the spatial distribution of strain inside a polycrystalline sample from the change in the neutron transmission spectra near a Bragg edge~\cite{abbey2009feasibility,abbey2012neutron,abbey2012reconstruction,lionheart2015diffraction,gregg2017bragg}. Given ideal conditions and a uniformly strained material, the Bragg edge can be modelled as a Heaviside function multiplied by a linear function of wavelength~\cite{lionheart2018histogram}. The result of this uniform strain is to shift the relative position (mean) of the Bragg edge with respect to that for a sample without strain present. However, the mean cannot provide sufficient information to resolve the strain distribution along the ray path~\cite{lionheart2015diffraction}, \emph{i.e.} there are infinitely many distributions of the strain fields along the beam path which will produce the same mean. This problem is related to a non-trivial null-space of the longitudinal ray transform, which gives a mathematical foundation for Bragg edge strain tomography~\cite{lionheart2015diffraction}, \emph{i.e.}, the mean measurements do not uniquely determine strain tensor fields. To overcome this problem, tomographic data can be combined with equilibrium equations of elasticity using a finite element approach to find the strain~\cite{gregg2017bragg}. Alternatively, Lionheart~\cite{lionheart2018histogram} observed that an experimentally measured Bragg edge is representative of the cumulative strain histogram along a neutron ray within the material. Hence, differentiation of the Bragg edge will theoretically return the histogram of strain, \emph{i.e.}, the distribution of strain components collinear with the ray disctretized into bins. The shape of the histogram is the convolution of the histogram for the unstrained case with the histogram of the relevant component of strain along the beam, and the second moment of the deconvolved histogram is the ray transform of the symmetric second tensor power of the strain. Hence, the histogram longitudinal ray transform~\cite{lionheart2018histogram} can be used to reconstruct the strain tensor in every voxel. The proposed theoretical method relies on our ability to measure the second moment of the strain distribution in transmission (projection) data which has not yet been demonstrated in practice. 

\begin{figure*} 
	\centering
	\includegraphics[width=\linewidth]{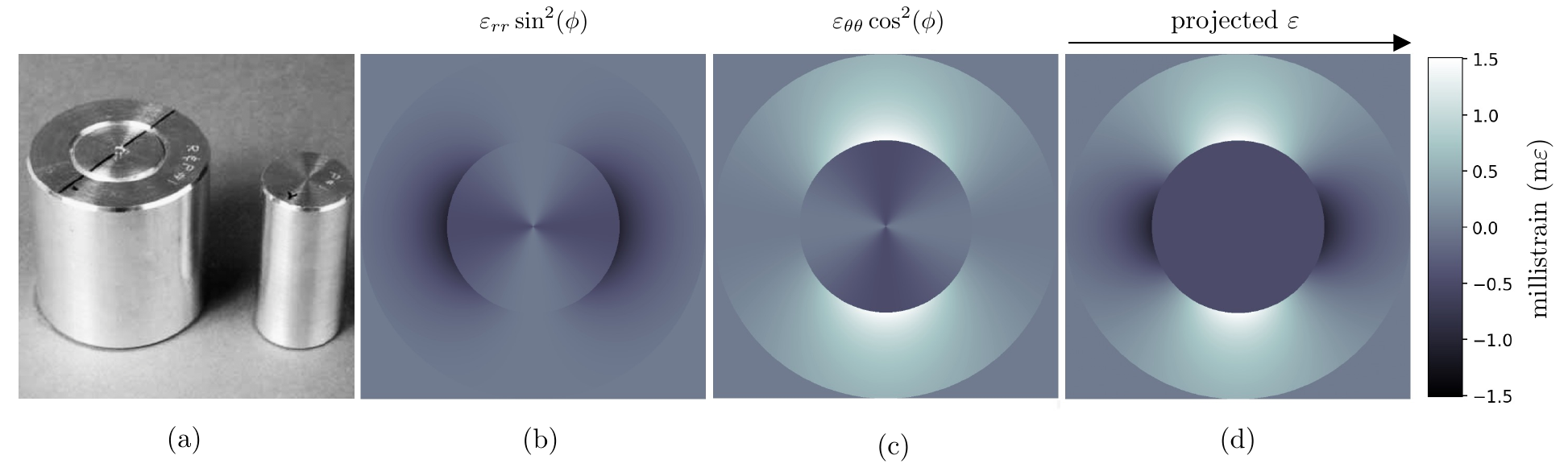}
	\caption{\label{fig:vamas} a) VAMAS round robin shrink fitted aluminium ring-and-plug and plug test samples (figure taken from the VAMAS report~\cite{webster2000neutron}). b) $\epsilon_{rr}$ component of the strain tensor scaled by $\sin^2 \phi$.   c) $\epsilon_{\theta \theta}$ component of the strain tensor scaled by $\cos^2 \phi$. d) Plot of the expected strain ($\epsilon = \epsilon_{\theta \theta} \cos^2 \phi + \epsilon_{rr} \sin^2 \phi$) within the strained sample for a ray path indicated by the arrow above the figure.}
\end{figure*}

In this paper, we demonstrate that the second moment, the variance about zero, of strain in the ray direction can be captured experimentally. We present an analysis of two reference samples (Fig.~\ref{fig:vamas}, a) manufactured within the Versailles Project on Advanced Materials and Standards (VAMAS)~\cite{webster2001polycrystalline}. The first sample is a shrink-fit aluminium alloy assembly of ring and plug (henceforth the \emph{strained} sample). The ring and the plug have outer diameters of 50~mm and 25~mm, respectively. The second sample is an unstained plug of the same diameter (henceforth the \emph{strain-free} sample). Both samples were manufactured under well controlled conditions, are of weak crystallographic texture and low residual stress prior to assembly. In addition they have been extremely well characterised in a global round-robin study~\cite{daymond2002analysis}.

In order to compare experimental measurements with theoretical predictions, we briefly recall some details about the expected strain in the strained sample. The axial stress of the plug, $\sigma_{zz}^{p}$, the ring $\sigma_{zz}^{r}$, and the interface pressure, $P$, have been determined in a series of neutron diffraction strain experiments~\cite{daymond2002analysis} and have been found to be -15~MPa, 5~MPa and 48~MPa, respectively. The values of Poisson's ratio, $\nu$, and the Young's modulus, $E$, of the material were taken to be 0.33 and 68~GPa, respectively. The radial and hoop strain can be obtained via solving the governing equations of linear elastic theory~\cite{boin2011developments}. Assuming that the axis of the cylinder is perpendicular to the direction of travel of the sufficiently parallel neutron beam, the strain along the ray path, $\epsilon$, is related to the radial ($\epsilon_{rr}$) and hoop ($\epsilon_{\theta \theta}$) components of strain via $\epsilon = \epsilon_{rr} \sin^2 \phi + \epsilon_{\theta \theta} \cos^2 \phi$, where $\phi$ is the angle anticlockwise from the neutron direction of travel. We will refer to the resulting distribution of strain as the \emph{projected strain}. We discretise the analytical strain map onto the experimental detector grid and sample from this array to calculate the first and the second moments of the distribution along the ray path. Fig~\ref{fig:vamas}, b-d shows the individual contour maps of the two calculated components of strain and their sum.

\begin{figure*}
	\includegraphics[width=\linewidth,keepaspectratio]{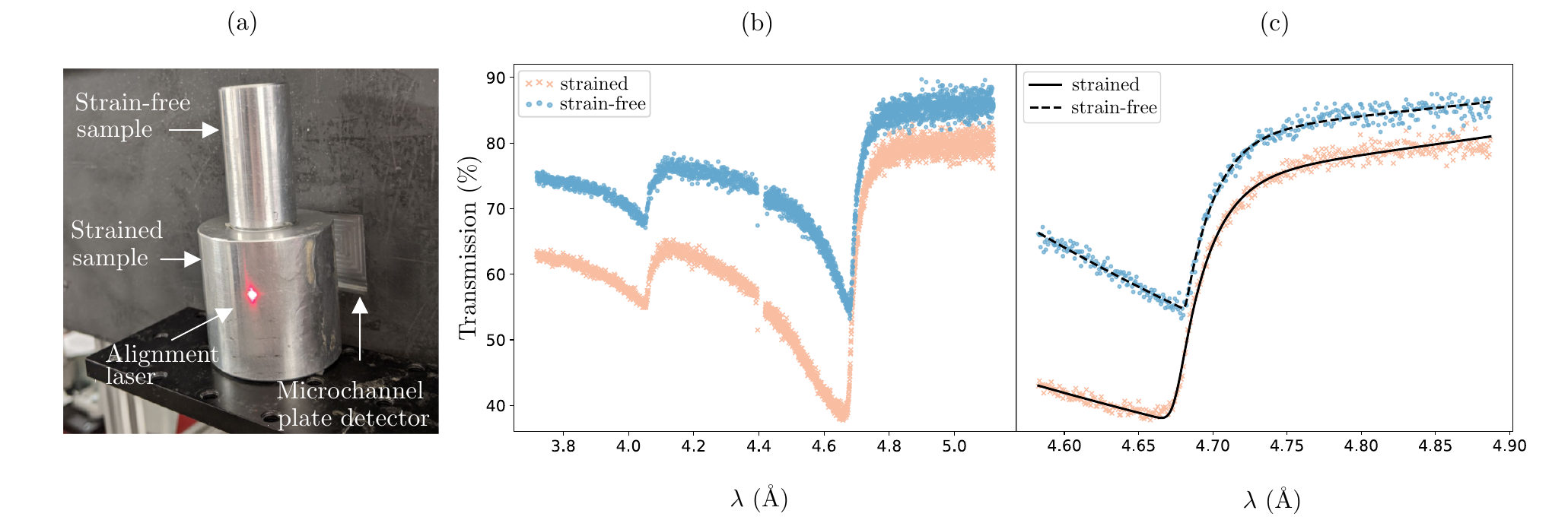}
	\caption{\label{fig:daq} a) Scheme illustrating experimental data acquisition. b) Plot showing the transmission of neutrons (\%) in a single macro-pixel as a function of wavelength for both samples. Both curves are plotted with the same vertical axis. As samples have different diameters and have been positioned slightly differently (translationally), a pixel with the same index in both transmission images will correspond to a different penetration length through the material, hence, intensity. Therefore there is a vertical offset between the two plotted curves. A gap in the recorded spectra is caused by detector readout between two shutter intervals. c) Measured Bragg edge in a single macro-pixel overlaid with the fitted model function.} 
\end{figure*}

The sample was measured~\cite{lionheart2019tof} at the Imaging and Materials Science \& Engineering (IMAT) beamline operating at the ISIS spallation neutron source (Rutherford Appleton Laboratory, UK)~\cite{burca2013modelling,kockelmann2018time}. At a pulsed neutron source, the wavelengths of the detected neutrons are calculated from their time of flight by 
\begin{equation}
\lambda = \frac{h (T+\Delta T_0)}{m L}
\end{equation} 
where $\lambda$ is the neutron wavelength (in meters), $h$ is Planck's constant, $T$ is the neutron time of flight (in seconds), $\Delta T_0$ is the time offset of the source trigger received by the data processing electronics (in seconds), $m$ is the neutron mass (in kilograms), and $L$ is the flight path from source to the detector (in meters). The MCP detector~\cite{tremsin2011high,tremsin2012high} used for the experiment was configured to record 2897 wavelength channels between 3.12~\AA~and 5.12~\AA~giving access to lattice planes from 1.56~\AA~to 2.56~\AA~in $d$-spacing, which for aluminum are the 111 and 200 lattice planes. To reduce the undesirable effect of counts loss~\cite{tremsin2014optimization}, two shutter intervals were set in the ToF (wavelength) domain with resolution $7.21 \cdot 10^{-4}$~\AA \, and $3.60 \cdot 10^{-4}$~\AA, respectively. The MCP detector has $512 \times 512$  pixels, 0.055~mm pixel size, giving a field of view of approximately $28 \times 28~\mathrm{mm}^2$. A visible laser beam was used to align the cylinder axis of the sample with respect to the vertical edge of the detector and to ensure that the plug, the ring and their interface are in the field-of-view (Fig.~\ref{fig:daq}, a). Subsequently, the strain-free reference sample was aligned and centered vertically and measured.

Individual projections of samples and a single normalisation image were measured using 4 hours long exposures. Flat-field and MCP detector related corrections~\cite{tremsin2014optimization} were adapted from BEAn~\cite{liptak2019developments} and applied to the projections. As the strained sample is axially symmetric, of weak crystallographic texture and only a radial-hoop internal stress exists, we assume that the strain does not vary along the cylinder axis. Consequently, we sum over each vertical column of pixels to improve the signal-to-noise ratio. Such  aggregated pixels are commonly referred to as \emph{macro-pixels}.

Given the measured wavelength range, two distinct Bragg edges were present in the acquired spectra, $\lambda \approx 4.0$~\AA~(200 lattice planes), and $\lambda \approx 4.7$~\AA~(111 lattice planes), with the latter one more pronounced and also sampled with higher wavelength resolution (Fig.~\ref{fig:daq}, b). Therefore we performed analysis only for the latter edge. To model the transmission spectra around the Bragg edge we used the Santisteban function~\cite{santisteban2001time}
\begin{equation} \label{eq:model_function}
\operatorname{Tr}(\Lambda, \psi) = \exp(-(a_0 + b_0 \Lambda))( 1- \exp(-(a_1 + b_1\Lambda))) B(\Lambda)
\end{equation}
\noindent
where $\Lambda$ is the experimentally acquired transmission signal measured in \AA~and $\psi = (a_0, b_0, a_1, b_1)$ is a vector of the model parameters.

Here, $a_0,b_0$ and $a_1, b_1$ describe the exponential attenuation to the right (tail) and to the left (pedestal) of the Bragg-edge, respectively, and $B(\Lambda)$ is given by
\begin{equation*}
	B\left(\Lambda \right) = \frac{1}{2} \Big[ \operatorname{erfc} \left( -\frac{\Lambda - \lambda_{hkl}}{\sqrt{2} \sigma} \right) -\exp \left(
	-\frac{\Lambda - \lambda_{hkl}}{\tau}+\frac{\sigma^{2}}{2 \tau^{2}}
	\right) \times \operatorname{erfc} \left( -\frac{\Lambda - \lambda_{hkl}}{\sqrt{2}\sigma}+\frac{\sigma}{\tau} \right) \Big].
\end{equation*}
where $\lambda_{hkl}$ is the position of the Bragg-edge, $\tau$ is the moderator decay constant, and $\sigma$ is the Gaussian broadening due to the sample and instrument. See the supplementary material for derivation of the model.

The model~\cite{santisteban2001time} was not explicitly derived to account for strain but rather to model blur in the wavelength dimension due to the stochastic nature of neutron moderation and the geometric effects of the beamline~\cite{suortti1979voigt,kropff1982bragg}. Nevertheless the sensitivity of the model to strain has been demonstrated in several studies~\cite{santisteban2002strain,woracek2011neutron,tremsin2012strain,tremsin2014strain}. 

We used the non-linear least square fitting (Levenberg?Marquardt algorithm) to fit the model function (eq.~\ref{eq:model_function}) and estimate parameters. To avoid the local minimum problem common to the non-linear fitting, we employ a three stage fitting process~\cite{santisteban2001time}. An example of a measured Bragg edge overlaid with the fitted model function is shown in Fig.~\ref{fig:daq}, c.

The \emph{first moment} (mean) of the projected strain is given by~\cite{hendriks2020bayesian}:
\begin{equation}
\langle \epsilon \rangle = \frac{\lambda^{s}_{hkl}-\lambda^{0}_{hkl}}{\lambda^{0}_{hkl}}
\end{equation}
where $\lambda^{0}_{hkl}$ and $\lambda^{s}_{hkl}$ are the position of Bragg edge for the strain-free sample and the strained sample, respectively.

Our  strategy for the measurement of the \emph{second moment} (variance about zero) is as follows. The value of the moderator decay constant, $\tau$, is a function of the geometry and temperature of the moderator used in the experiment~\cite{liptak2019developments}; as these parameters remained approximately constant within the experiment, $\tau$ is expected to be constant. Parameter $\sigma$ is a function of width of the initial pulse from the moderator and sample-related broadening~\cite{ramadhan2019characterization}. As the shape of the pulse is expected to be repeatable and uniform in the spatial dimension, any spatial change in $\sigma$ can be attributed to the change in variance of strain in the beam direction. Although $\sigma$ in eq.~\ref{eq:model_function} captures broadening of the Bragg edge, we need to establish a relationship between an instrument response and the variance of strain along a beam direction. Assuming a linear relationship, the model of measurements is given by $\mathrm{y} = m \mathrm{x}+c$, where $\mathrm{x} = [x_0, x_1, \ldots, x_{j-1} ]$ is a vector of the theoretically predicted variance of strain in the beam direction at detector macro-pixel  $j$ between 0 and 511 and $\mathrm{y} = [\sigma_0, \sigma_1, \ldots, \sigma_{j-1}]$ is experimentally measured $\sigma$ at each macro-pixel $j$. We use linear regression to define parameters $m$ and $c$. Obviously, this simple proof-of-concept measurement model cannot substitute a proper instrument scale calibration necessary to establish this tomographic measurement technique.

To support our findings, we perform uncertainty quantification based on Bayesian interference~\cite{jaynes2003probability}. In the Bayesian framework, the measurement model is represented as a joint probability distribution of unknown parameters $\boldsymbol{\eta}$ and observations~$\textbf{Y}$
\begin{equation}\label{bayesinfer}
\pi(\boldsymbol{\eta}|\textbf{Y}) = \frac{\pi(\textbf{Y}|\boldsymbol{\eta})\pi(\boldsymbol{\eta)}}{\pi(\textbf{Y})},
\end{equation}
where $\pi(\textbf{Y}|\boldsymbol{\eta})$ is the \emph{likelihood} function of $\boldsymbol{\eta}$, \emph{i.e.}, the predictive distribution of $\textbf{Y}$, given $\boldsymbol{\eta}$. The \emph{prior} distribution $\pi(\boldsymbol{\eta)}$ encodes the prior knowledge and model assumptions. The \emph{model evidence} $\pi(\textbf{Y})$ maps the likelihood, prior and observations to a single value that describes the probability of observation. Finally, $\pi(\boldsymbol{\eta}|\textbf{Y})$ is the \emph{posterior} probability: the probability of $\boldsymbol{\eta}$ after $\textbf{Y}$ is observed.

The mean of the likelihood is given by the parametric model for each data point. Let $j$ between 0 and 511 denote the position of a column of pixels and $\left[\operatorname{\textbf{Y}}_j\right]_i = \operatorname{Y}(\lambda_i)_j$ be the mean measured transmission for wavelength bin $i$ in pixel column $j$. We model the transmission error over a macro-pixel as additive Gaussian noise with zero mean and a variance that is linearly dependent on the transmission~\cite{hendriks2020bayesian}. Then,
\begin{equation}
\operatorname{Y}(\lambda_i)_j = \operatorname{Tr}(\lambda_i| \boldsymbol{\psi}_j) + \xi(\omega| \lambda_i, j),
\end{equation}
where $\xi(\omega| \lambda_i, j)$ is a Gaussian random variable  
\begin{equation}
\xi(\omega| \lambda_i, j) \sim \mathcal{N}(0, s(\lambda_i)_j^2),   
\end{equation}
and $s(\lambda_i)_j^2$ is the unbiased estimate of the sample variance. Then, the likelihood $\pi(\textbf{Y}|\boldsymbol{\eta})$ is given by 
\begin{align}
\pi(\operatorname{Y}(\lambda_i)_j|\boldsymbol{\psi}_j) &= \pi(\xi(\omega| \lambda_i, j) = \operatorname{Y}(\lambda_i)_j - \operatorname{Tr}(\lambda_i| \boldsymbol{\psi}_j))\\ 
&\propto \exp\left(- \frac{1}{2}||\operatorname{Y}(\lambda_i)_j - \operatorname{Tr}(\lambda_i| \boldsymbol{\psi}_j)||_{\Sigma}^2 \right),
\end{align}
where $||\cdot||^2_{\Sigma}$ is the covariance-weighted norm. Finally, we converge to
\begin{equation}
\pi(\operatorname{Y}(\lambda_i)_j|\boldsymbol{\psi}_j) = \exp\left(-\sum_{\lambda_i}\frac{1}{2 s(\lambda_i)}\Big(\operatorname{Y}(\lambda_i)-\operatorname{Tr}(\lambda_i| \boldsymbol{\psi}_j)\Big)^2\right). 
\end{equation}

The prior distributions on the parameters of the model ${\psi}_j$ are assumed to be weakly informative (wide peak) Gaussian's centered at the best estimates obtained from the Levenberg-Marquardt fit for each parameter. We further use the Hamiltonian Monte Carlo (HMC)~\cite{betancourt2017conceptual} method to sample from the posterior distribution. Bayesian interference and HMC are implemented using the Python wrapper PySTAN for the probabilistic programming framework STAN. 

\begin{figure}
	\centering
	\includegraphics[width =0.48\textwidth,keepaspectratio]{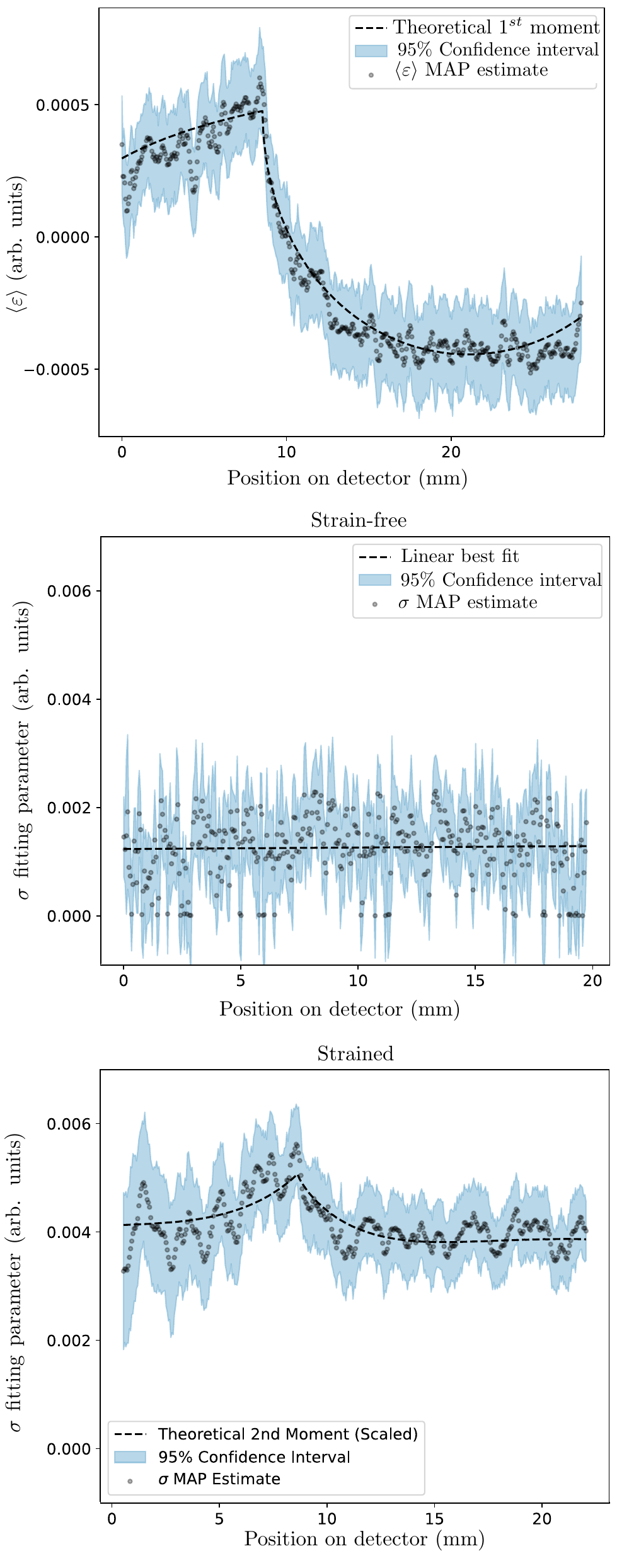}
	\caption{\label{fig:results} Variation in the second moments as a function of detector pixel obtained via HMC Bayesian inference. MAP estimate refers to the position of the maximum posterior density and the 95\% is found from $\pm 2$ standard deviations of the posterior distribution. }
	\label{fig:}
\end{figure}

Fig.~\ref{fig:results} compares a maximum a posteriori probability (MAP) estimation of the mean and second moment obtained from the  experimental data. Overlaid we plot a confidence interval of two standard deviations of the distribution. For the strain-free case, parameter $\sigma$ is expected to be constant but greater than 0 as $\sigma$ also models blur in the wavelength dimension. Therefore the theoretical predictions are given by the best linear fit to estimated data. Theoretical predictions for the strained sample are given by our calculations in fig.~\ref{fig:vamas}, d which were scaled linearly to best match data. It can clearly be seen that both mean and variance are within the uncertainty interval for both samples and and the main trends are captured. However there is strong noise present in all estimated parameters and for some data points the MAP estimate of the second moment is 0 and the 95\% confidence interval includes negative values. There are several reasons for the observed behaviours.

\begin{figure*}
	\centering
	\includegraphics[width=\linewidth,keepaspectratio]{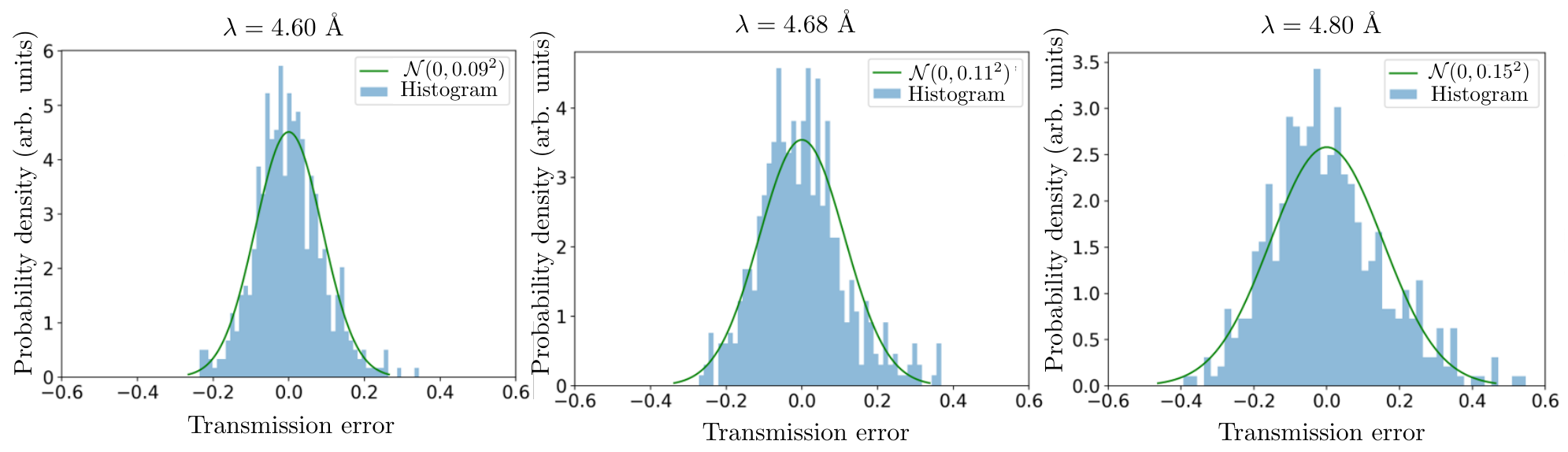}
	\caption{\label{fig:distributions} Histograms showing the distribution of the error for different values of transmission for the central column of the detector pixels. A Gaussian probability density function has been fitted to the data (solid line).}
\end{figure*}

Following Hendriks et al.~\cite{hendriks2020bayesian}, we assumed Gaussian noise in the measured transmission data. In fig.~\ref{fig:distributions} we show the distribution of error in some representative macro-pixels overlaid with the fitted Gaussian probability density function. While the distributions have a clear bell-shape, they are also skewed towards negative values. Conducting a combined D?Agostino and Pearson?s omnibus test~\cite{d1971omnibus} with a significance level of $\alpha = 0.001$ showed that of the 741,376 distributions considered 690,688 have enough evidence to reject the hypothesis that the data was drawn from a Gaussian distribution. The reason for this skew might be the overlap correction~\cite{tremsin2014optimization} used to compensate for counts loss. The correction relies on Poisson statistics and the weighting factor for each wavelength bin is calculated based on values in shorter wavelength bins introducing inter-bin correlations and potentially a skew in the data. 

\begin{figure*}
	\centering
	\includegraphics[width=\linewidth,keepaspectratio]{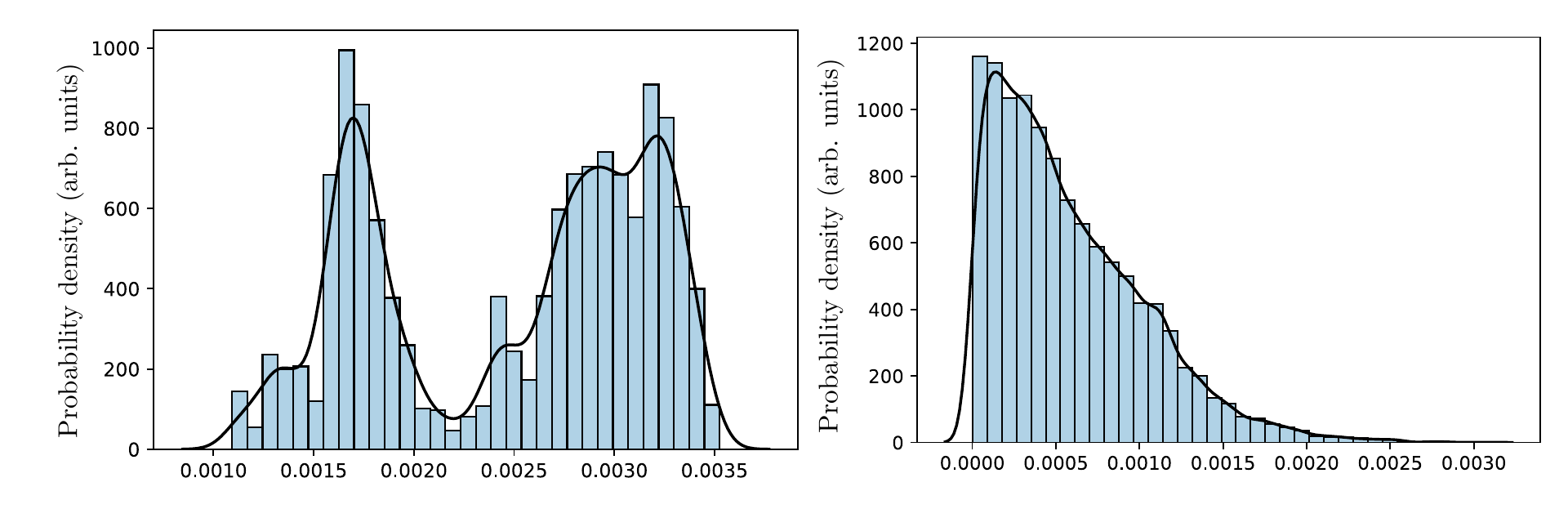}
	\caption{\label{fig:posterior} Histograms of the posterior density of the $\sigma$ parameter for detector column 110 of the strain-free sample (left) and for detector column 90 of the strained sample (right) shown; the solid line shows the kernel density estimate plot for the same distribution.}
\end{figure*}

The Santisteban model~\cite{santisteban2001time} was not designed to account for strain and the parameter $\sigma$, which was used in this study as a measure of strain variance, does not have any physical meaning in the model. Secondly, the model assumes the Gaussian distribution of strain. In fig.~\ref{fig:posterior} we show a posterior distribution of $\sigma$ for both samples. For the strained sample we chose a data point where the fitting resulted in $\sigma = 0$. The posterior distribution is concentrated at $\sigma = 0$, consistent with the least-square fit. For the strain-free sample, the posterior distribution is multimodal with two pronounced peaks. Both distributions highlight inadequacy of the Santisteban model for uniquely identifying the second moment of the strain distribution. Therefore, more research is needed to have more accurate physical Bragg edge models for strain measurements. 

While the demonstration in this paper is limited to a texture-free sample, the strain reconstruction approach~\cite{lionheart2018histogram} is applicable to a more general case of a textured sample. In the general case, we need to decouple various crystallographic information encoded in the measured Bragg edges. The actual shape of a Bragg edge in transmitted neutron beam reflects the crystal structure averaged along the beam path. With the current Bragg edge model we explicitly assume that the strain distribution along the beam path is Gaussian. Secondly, crystallographic texture (significant preferred orientation of crystallites) might not be properly handled by the current Bragg edge model. The texture affects the number of crystallites for which the backscattering condition is fulfilled~\cite{Vogel_2000,boin2011validation}. Hence, it will affect the edge?s pedestal, tail, and height. The current Bragg edge model was not designed to account for these effects and might not fit the data in the presence of significant texture. Hence, either a more accurate physical modelling of a Bragg edge is needed to decouple the effects of texture and strain on the Bragg edge, or direct derivation and deconvolution of Bragg edge data can be used. With the current measurement setup, the later method is not feasible due to low wavelength resolution near the edge.

To conclude, we have demonstrated that the second-order moment of the strain distribution can be obtained experimentally. The theoretically predicted first and second moments are covered by the 95\% confidence interval estimated through Bayesian inference. However we found out that Gaussian nature of the transmission error could only be established with relatively low confidence. Further work in this area should seek to improve our confidence in the choice of likelihood. Furthermore, the posterior distribution shows direct evidence that the semi-empirical Santisteban model is inadequate for uniquely extracting higher order moments in general. Therefore a model that explicitly accounts for the moments of the strain distribution and for texture effects in the material is needed. Despite the limitations of the current study, our findings pave the way for neutron strain tomography. The task of more accurate Bragg edge modeling, calibration and uncertainty quantification is an opportunity for future research.

See the supplementary material for derivation of the Santisteban model.

\section*{\label{sec:acknowledgment}Acknowledgment}

Authors would like to thank Dr S. Cotter, and Dr J. Hendriks for their input and ideas at various stages in this project.

\section*{\label{sec:data_availability}Data availability statement}

The data that support the findings of this study will be openly available following an embargo at the following URL/DOI: \href{https://doi.org/10.5286/ISIS.E.RB1920056}{10.5286/ISIS.E.RB1920056}~\cite{lionheart2019tof}. Data will be available from 27 November 2022. Until the embargo period expires, data are available from the corresponding author upon reasonable request. 

\section*{\label{Funding}Funding}
\sloppy
This work was funded by EPSRC grants ``A Reconstruction Toolkit for Multichannel CT'' (EP/P02226X/1) and ``Rich Nonlinear Tomography for advanced materials'' (EP/V007742/1). We gratefully acknowledge beamtime RB1920056 (URL/DOI: \href{https://doi.org/10.5286/ISIS.E.RB1920056}{10.5286/ISIS.E.RB1920056})~\cite{lionheart2019tof} at the IMAT Beamline of the ISIS Neutron and Muon Source, Harwell, UK. K F acknowledges the support of the  EPSRC grant ``EPSRC Centre for Doctoral Training in Agri-Food Robotics'' (EP/S023917/1) in the final stages of this work. E A was partially funded by the Federal Ministry of Education and Research (BMBF) and the Baden-Württemberg Ministry of Science as part of the Excellence Strategy of the German Federal and State Governments. W R B L acknowledges support from a Royal Society Wolfson Research Merit Award.

\printbibliography

\newpage
\input{supp.tex}
\end{document}

%% file: supp.tex
\setcounter{figure}{0}
\renewcommand{\figurename}{Fig.}
\renewcommand{\thefigure}{S\arabic{figure}}

\setcounter{equation}{0}
\renewcommand{\theequation}{S\arabic{equation}}

\section*{Supplementary material}

\begin{refsection}

Here we present the model used to fit experimental data, and infer material characteristics, used in the main paper. The predominant model for Bragg edge fitting is presented by Santisteban~\cite{Santisteban:2001}; the model uses the Kropff `resolution function' as a basis~\cite{Kropff1982}. Although the above-cited papers outline the procedure by which the parametric fitting function can be obtained, we note they lack any detailed derivations. In the following sections, we will provide the reader with a derivation of the Kropff resolution function and the Santisteban parametric model.

\subsection*{The Resolution Function}\label{Resolutionfn}
\begin{figure}[h]
	\begin{subfigure}[t]{.48\textwidth}
		\centering
		\includegraphics[width=1\linewidth]{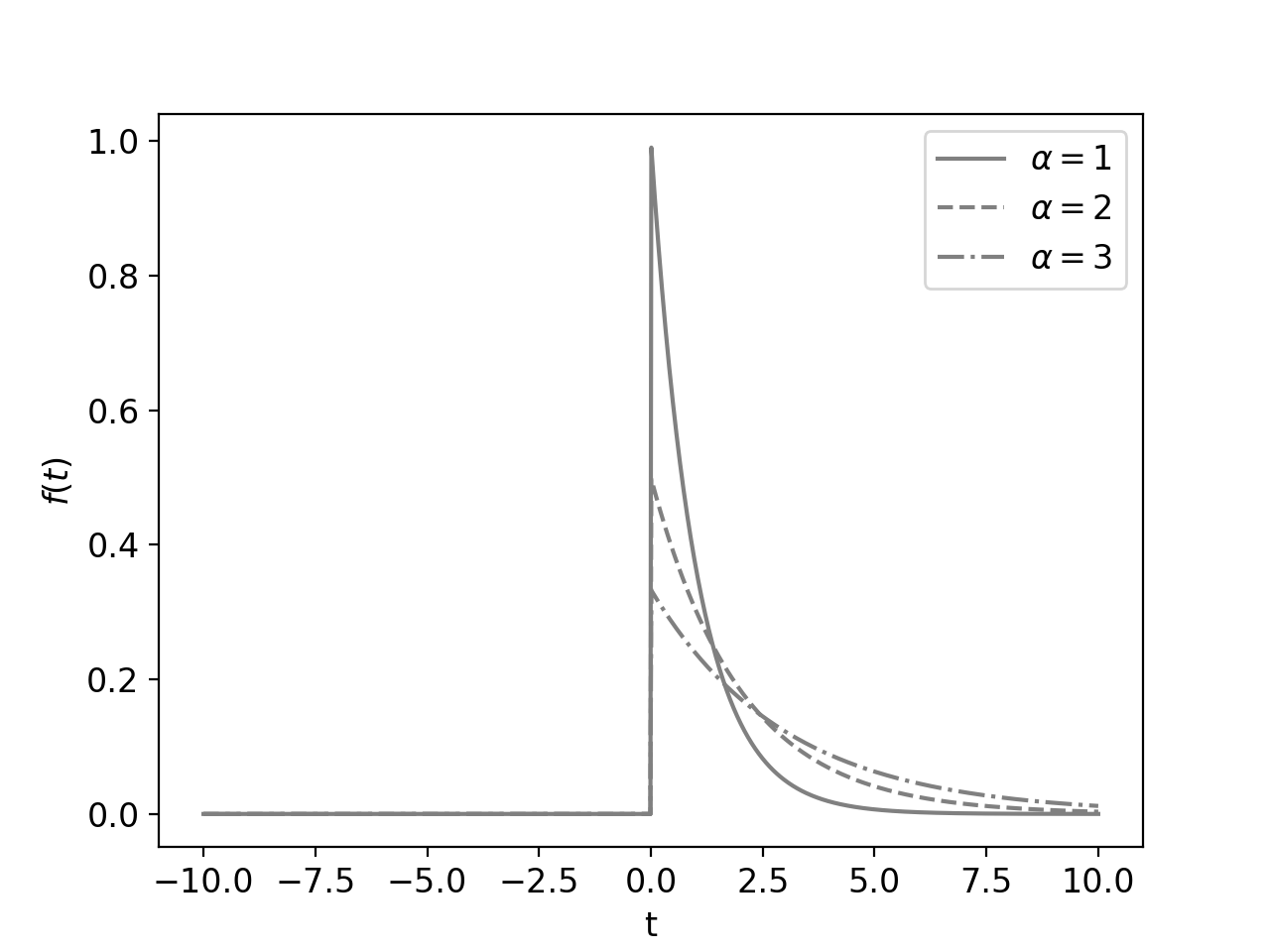}
		\caption{Source emission time distribution function, given by $f(t)$, for $t_0 =0$ and a range of $\alpha$.}
		\label{fig:fx}
	\end{subfigure}%
	\hfill
	\begin{subfigure}[t]{.48\textwidth}
		\centering
		\includegraphics[width=1\linewidth]{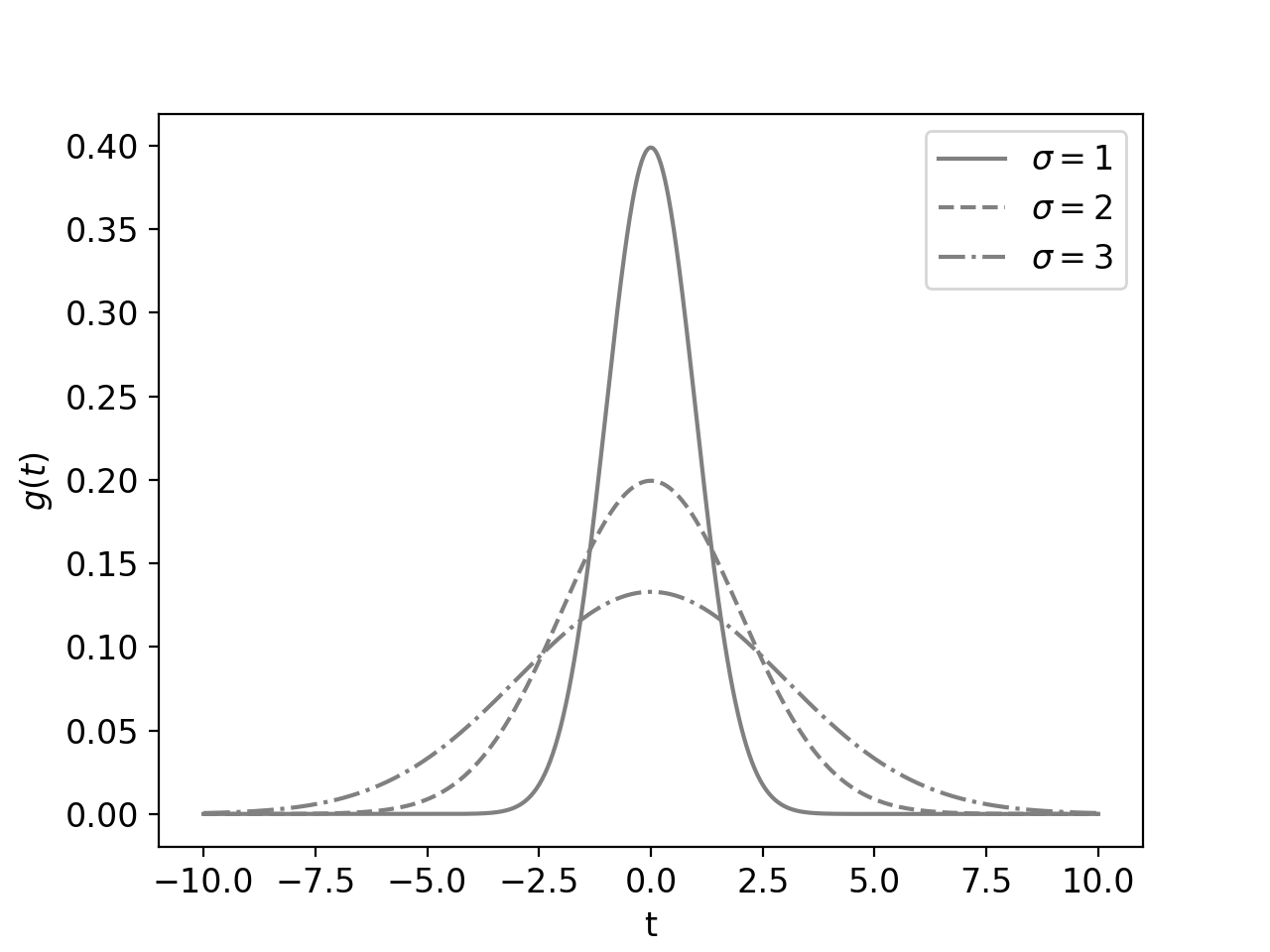}
		\caption{Geometric effects described by a Gaussian distribution, $g(t)$, for $t_0 = 0$ and a range of $\sigma$ values.}
		\label{fig:gx}
	\end{subfigure}
	\caption{Examples of the function $g(x)$ and $f(x)$ for a range of $\alpha$ and $\sigma$.}
	\label{fig:}
\end{figure}

The resolution function models the uncertainty in the time and position of neutrons leaving the moderator. To derive the model, we will work in the time domain (an equivalent spatial definition can be found via inference of the neutrons wavelength using the \textit{de Broglie} relation). Following the argumentation from~\cite{Kropff1982}, we assume that the uncertainty in resolution is controlled by two phenomena: 
\begin{enumerate}
	\item {The stochastic nature of neutron moderation is modelled by a source emission time distribution function, $f(t)$, given by a product of a Heaviside function, $\mathcal{H}(t)$, and a decaying exponential, with both functions centered at $t_0$ (fig.~\ref{fig:fx}),
		\begin{equation}\label{f(x)}
		f(t) = \frac{\mathcal{H}(t-t_0)}{\alpha} \exp\left(-\frac{t-t_0}{\alpha}\right),
		\end{equation}
		where 
		\begin{equation}
		\mathcal{H}(x) =
		\begin{cases}
		1, & x\ge0, \\
		0, & x<0.
		\end{cases}
		\end{equation}}
	\item {The geometric effects of the beamline have empirically been shown to distort Bragg-lines into Gaussian shaped curve \cite{Suortti1979}. The geometric effects are modelled by a normalised Gaussian function $g(t)$ centred at $t_0$ (fig.~\ref{fig:gx}), 
		\begin{equation}\label{g(x)}
		g(t) = \frac{1}{\sqrt{2\pi}\sigma}\exp\left( -\frac{(t-t_0)^2}{2\sigma^2}\right).
		\end{equation}}
\end{enumerate}

The combined effect of the source time emission and the geometric effects of the beamline is given by the convolution of functions $f(t)$ and $g(t)$. Let $t-t_0 = \tau$, then the convolution is given by 
\begin{align}
\begin{split}
I(\tau) = \int_{-\infty}^{\infty} f(x) g(\tau - x) dx &= \int_{-\infty} ^{\infty}  \frac{\mathcal{H}(x)}{\sqrt{2\pi}\sigma\alpha} \exp\left(-\frac{x}{\alpha}\right)\exp\left( -\frac{(\tau-x)^2}{2\sigma^2}\right) dx\\
&= \int_0^\infty \frac{1}{\sqrt{2\pi}\sigma\alpha} \exp\left(-\frac{x}{\alpha}\right)\exp\left( -\frac{(\tau-x)^2}{2\sigma^2}\right)dx\\
&= \int_0^\infty \frac{1}{\sqrt{2\pi}\sigma\alpha}\exp\left( -\frac{2\sigma^2 x +\alpha\tau^2 +\alpha x^2 -2\alpha \tau x}{2\alpha \sigma^2}\right)dx. \label{convintegral}
\end{split}
\end{align}
Let us subtract and add
\begin{equation}
\frac{2 \tau \alpha - \sigma^2}{2 \alpha^2}
\end{equation}
within the bracketed term in eq.~\ref{convintegral}. It follows
\begin{align}
\begin{split}
&\frac{2\sigma^2 x +\alpha\tau^2 +\alpha x^2 -2\alpha \tau x}{2\alpha \sigma^2} -\frac{2\tau\alpha -\sigma^2}{2\alpha^2} +\frac{2\tau \alpha -\sigma^2}{2\alpha^2} \\
& = \frac{\alpha^2 x^2 -2\alpha^2 \tau x +2\alpha\sigma^2 x +\alpha^2\tau^2 -2\tau\alpha\sigma^2 +\sigma^4}{2\alpha^2\sigma^2} +\frac{1}{\alpha}\left(\tau -\frac{\sigma^2}{2\alpha}\right) \\
& = \frac{\alpha^2 x^2 -2\alpha^2 \tau x +2\alpha\sigma^2 x -(\alpha\tau -\sigma^2)^2}{2\alpha^2\sigma^2} +\frac{1}{\alpha}\left(\tau -\frac{\sigma^2}{2\alpha}\right) \\
& = \frac{\left(\alpha x -(\alpha\tau -\sigma^2) \right)^2}{2\alpha^2\sigma^2} +\frac{1}{\alpha}\left(\tau -\frac{\sigma^2}{2\alpha}\right)\\
& = \left(\frac{1}{\sqrt{2}\sigma}\left(x-\left(\tau-\frac{\sigma^2}{\alpha}\right)\right)\right)^2 +\frac{1}{\alpha}\left(\tau -\frac{\sigma^2}{2\alpha}\right).
\label{rearrangement}
\end{split}
\end{align}
We now introduce 
\begin{equation}\label{sub}
z = \frac{1}{\sqrt{2}\sigma}\left(x-\left(\tau-\frac{\sigma^2}{\alpha}\right)\right).
\end{equation}
Treating $z$ as an implicit function of $x$ and differentiating we can find the required differential to be 
\begin{equation}\label{subder}
dx = \sqrt{2}\sigma dz.  
\end{equation}

The limits of integration, $x \to \infty$ and $x \to 0$, become $z \to \infty$ and $z \to \frac{-(\tau-\sigma^2/\alpha)}{\sqrt{2}\sigma}$, respectively. Utilizing eq.~\ref{rearrangement} and~\ref{sub} and changing the limits of integration result in
\begin{equation}
I(\tau) = \frac{\sqrt{2}\sigma}{\sqrt{2\pi}\sigma\alpha}\exp\left( - \left(\frac{\tau}{\alpha} -\frac{\sigma^2}{2\alpha^2} \right)\right)\int_{\frac{-(\tau-\sigma^2/\alpha)}{\sqrt{2}\sigma}}^\infty\exp\left( -z^2\right)dz.
\end{equation}

To converge to a closed-form solution of the convolution integral, we introduce the error function and the complementary error function. The error function, denoted by $\mathrm{erf}(z)$, is a complex function of a complex variable defined as
\begin{equation}
\mathrm{erf}(z) = \frac{2}{\sqrt{\pi}}\int _0 ^{z} \exp(-t^2)dt.
\end{equation}
The complementary error function is defined by 
\begin{equation}
\mathrm{erfc}(z) = 1 - \mathrm{erf}(z).
\end{equation}
Given the error function and its properties\footnote{The error function has the following properties:
	$$\mathrm{erf}(0) = 0 ~~~~~~ \mathrm{erf}(\infty) = 1 ~~~~~~ \mathrm{erf}(-x) = - \mathrm{erf}(x)$$
	$$\mathrm{erfc}(0) = 1 ~~~~~~ \mathrm{erfc}(\infty) = 1 ~~~~~~ \mathrm{erfc}(-x) =2 - \mathrm{erfc}(x)$$}, we can write  
\begin{equation}\label{Kropfff_Almost_done}
I(\tau) =  \frac{1}{2\alpha}\exp\left( - \left(\frac{\tau}{\alpha} -\frac{\sigma^2}{2\alpha^2} \right)\right)\mathrm{erfc}\left(\frac{-(\tau-\sigma^2/\alpha)}{\sqrt{2}\sigma}\right).
\end{equation}

\begin{figure}[h]
	\centering
	\includegraphics[width = 0.7\textwidth]{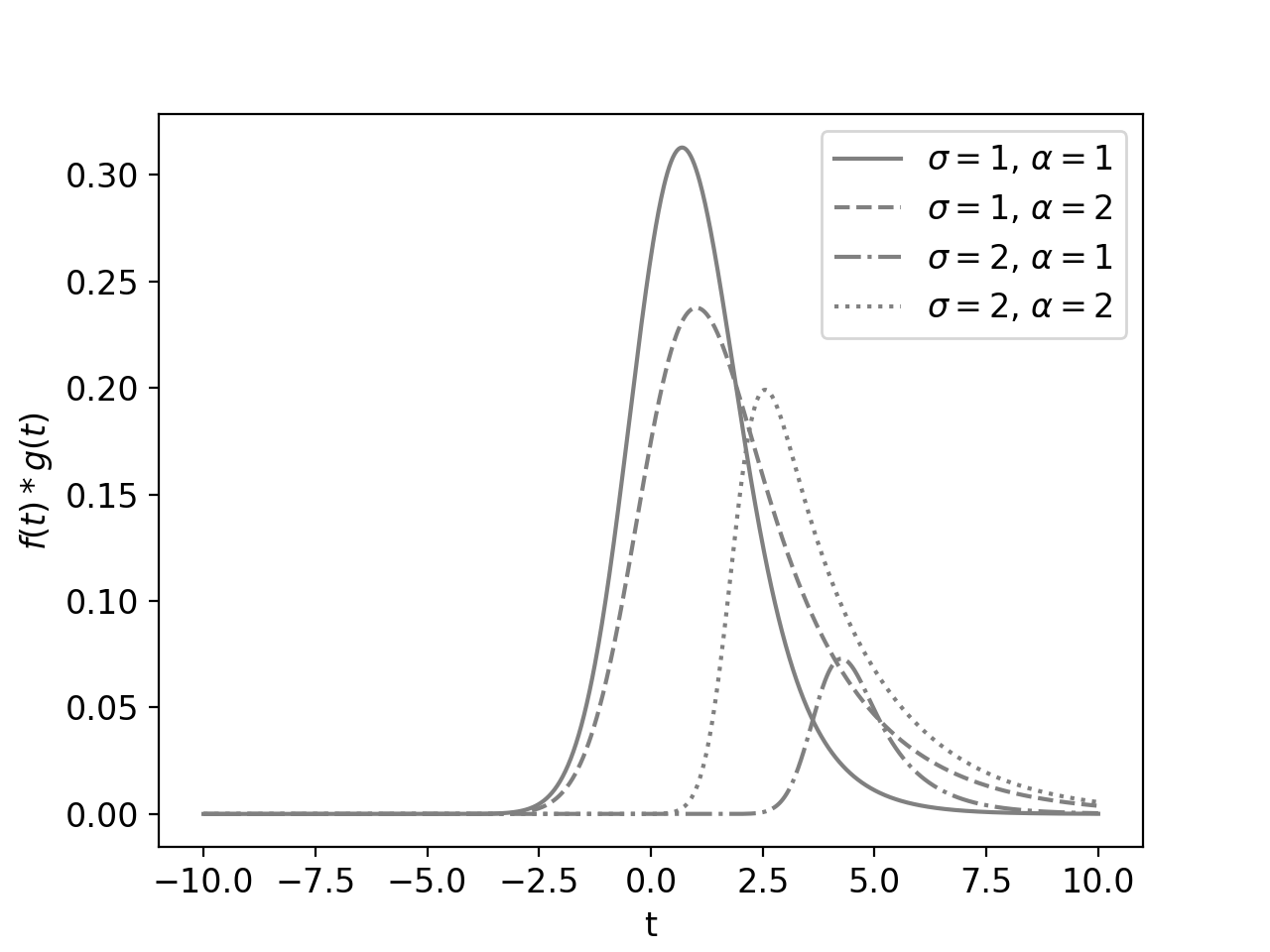}
	\caption{Plot of the convolution of the functions $f(t)$ and $g(t)$ for $t \in [-10,10]$ and $t_0 = 0$.}
	\label{fig:f(t)g(t)}
\end{figure}

Eq.~\ref{Kropfff_Almost_done} provides a closed form solution to the convolution in eq.~\ref{convintegral}; a plot for a range of $\alpha$'s and $\sigma$'s is given in fig.~\ref{fig:f(t)g(t)}.
\noindent
With some minor rearrangement we arrive at the form presented in the work by Kropff~\cite{Kropff1982};
by letting $$
u = \frac{1}{\sqrt{2}}\left(\frac{\tau}{\sigma} -\frac{\sigma}{\alpha}\right)
$$
and by noting that 
\begin{align}
\begin{split}
&\exp\left( - \left(\frac{\tau}{\alpha} -\frac{\sigma^2}{2\alpha^2}\right)\right)\\
&= \exp\left(-\frac{\sigma^2}{2\alpha^2}\right)\exp\left(-\left( \frac{\tau}{\alpha} -\frac{\sigma^2}{\alpha^2}\right)\right) \\
& =\exp\left(-\frac{\sigma^2}{2\alpha^2}\right)\exp\left(-\sqrt{2}\left(\frac{\sigma}{\alpha}\right)\frac{1}{\sqrt{2}}\left(\frac{\tau}{\sigma} -\frac{\sigma}{\alpha}\right)\right) 
\end{split}
\end{align}
we can write 
\begin{align}
I(\tau) &=  \frac{1}{2\alpha}\exp\left(-\frac{\sigma^2}{2\alpha^2}\right)\exp\left(-\sqrt{2}\left(\frac{\sigma}{\alpha}\right)\frac{1}{\sqrt{2}}\left(\frac{\tau}{\sigma} -\frac{\sigma}{\alpha}\right)\right)\mathrm{erfc}\left(-u\right) \nonumber\\ 
&= \frac{1}{2\alpha}\exp\left(-\frac{\sigma^2}{2\alpha^2}\right)\exp\left(-\sqrt{2}\left(\frac{\sigma}{\alpha}\right)\frac{1}{\sqrt{2}}\left(\frac{\tau}{\sigma} -\frac{\sigma}{\alpha}\right)\right) \left[1+\mathrm{erf}(u) \right].
\label{KropffEq}
\end{align}
Eq.~\ref{KropffEq} is equivalent to eq.~(6) presented in~\cite{Kropff1982}. Although this convolution-based approach, which splits the source emission time error, has been empirically shown to work very well, Kropff highlighted that this model represents an oversimplification of the underlying mechanics~\cite{Kropff1982}.

\subsection*{Neutron Transmission Model}\label{NTM}

Due to the nature of a pulsed neutron source, a broad range of neutron energies, and hence wavelengths, can be produced each time the spallation target is hit with high energy protons; we will refer to this distribution of neutron wavelengths as $S(\lambda)$, which is the expected number of neutrons produced of wavelength $\lambda$. Given this distribution, and the assumption that the number of neutrons is conserved between creation and detection, the total number of neutrons that should hit the detector between time $t$ and $t+\Delta t$ is 
\begin{equation}\label{N(t)Naive}
N(t) = \left(\int_0 ^{\infty} S(\lambda') d\lambda'\right)\Delta t.
\end{equation}
Eq.~\ref{N(t)Naive} however does not account for the resolution function~\ref{KropffEq}, the efficiency of the detector or the effect of placing objects in the beam path. Hence, eq.~\ref{N(t)Naive} becomes 
\begin{equation}\label{N(t)}
N(t) = \left(\int_0^{\infty} S(\lambda') T(\lambda') \epsilon(\lambda') R(\lambda') d\lambda' \right)\Delta t,
\end{equation}
\noindent
where $T(\lambda)$ is the transition probability, $\epsilon(\lambda)$ is the detector efficiency for wavelength $\lambda$ ($\epsilon(\lambda) = 1$ meaning that all neutrons transmitted are detected) and $R(\lambda)$ is the resolution function presented in the previous section. 

\begin{figure}[h]
	\centering
	\includegraphics[width = 0.75\textwidth]{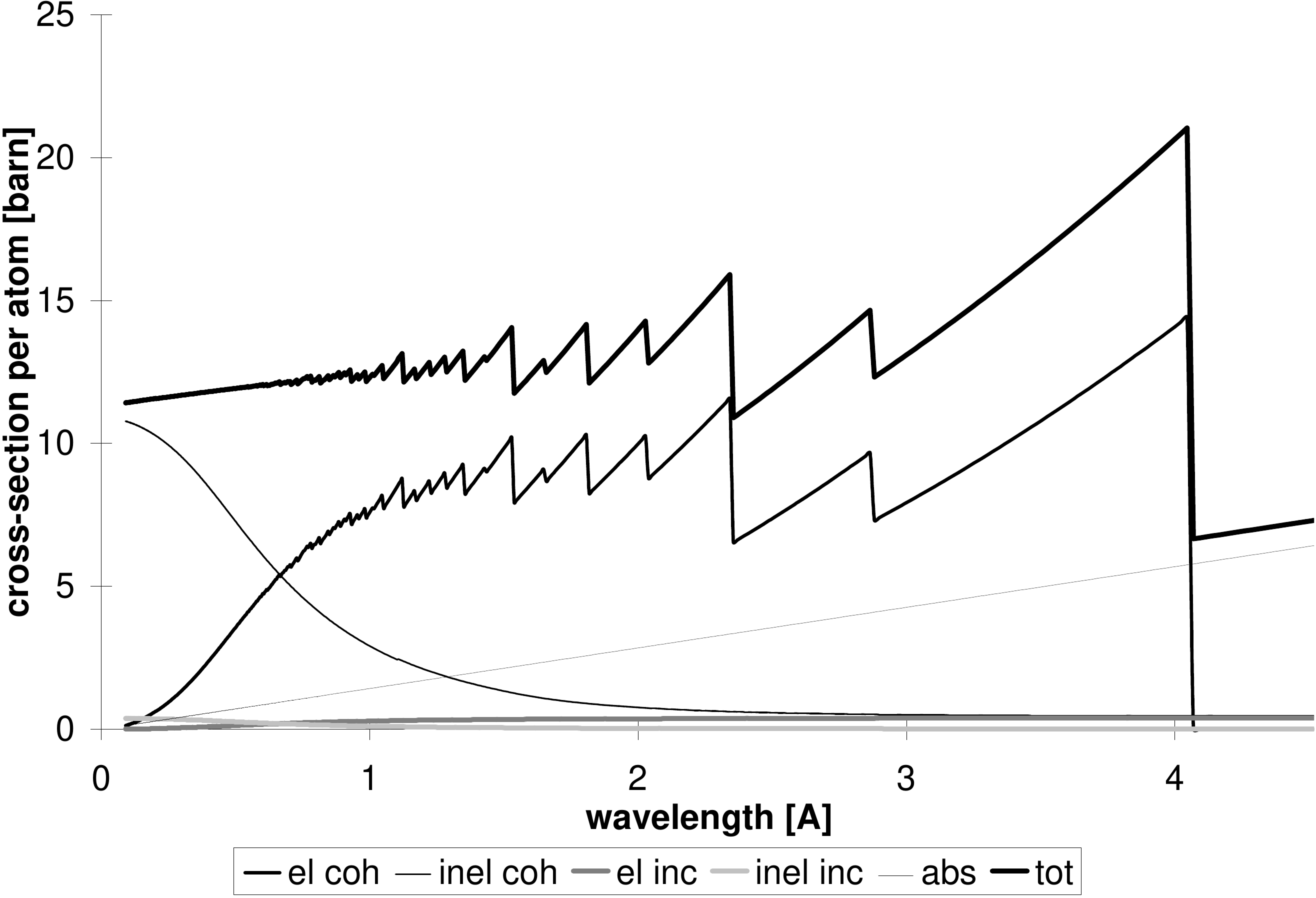}
	\caption{Cross-sections of $\alpha$-iron per atom calculated with the software BETMAn. Figure reproduced with permission from \cite{Vogel2000}.}
	\label{fig:cross-section}
\end{figure}

The physical fundamentals underlying nuclear interaction are described in numerous works; we will only scrape the surface of the massive field of neutron physics by covering fundamentals needed for the construction of the neutron transmission model. Neutrons are able to interact with matter is a variety of ways; a convenient way of describing each type of interaction, and there cumulative effect, is by the idea of a  \textit{cross-section}, $\sigma$.  A basic model for neutron transmission is then an exponential attenuation, 
\begin{equation}\label{TR(model)}
T(\lambda) = \exp(- nx\sigma_T(\lambda)),
\end{equation}
governed by the material thickness $x$, the number of scattering centers per unit volume $n$ and the total scattering cross-section $\sigma_T(\lambda)$ per scattering center. Restricting our consideration to only polycrystalline materials, the total scattering cross-section, $\sigma_T(\lambda)$, can further be broken down into several components,
\begin{equation}\label{eq:cross_section}
\sigma_T(\lambda) = \sigma_{el}^c(\lambda)+\sigma_{inel}^c(\lambda)+\sigma_{el}^{inc}(\lambda)+\sigma_{inel}^{inc}(\lambda),
\end{equation}
where subscripts $el$ and $inel$ stand for elastic and inelastic interactions and the superscripts $c$ and $inc$ stand for coherent and incoherent. A plot of typical collision cross-section is given in fig.~\ref{fig:cross-section}. For the materials employed in this study, the change in transmission near Bragg edges governed by the coherent elastic scattering, $\sigma_{el}^c$. Then, eq.~\ref{eq:cross_section} can be rewritten as
\begin{equation}\label{sigma_T}
\sigma_T (\lambda) = \sigma_0(\lambda) + \sigma_{hkl}(\lambda)\left(1 -\mathcal{H}(\lambda - 2d_{hkl})\right)
\end{equation}
where $\sigma_{hkl}$ accounts for the coherent elastic scattering contribution from the plane $hkl$, $\sigma_0$ accounts for all other contributions and $\mathcal{H}(\lambda - 2d_{hkl})$ is the Heaviside function centered at $\lambda_{hkl} = 2d_{hkl}$.\\

Substituting eq.~\ref{sigma_T} and \ref{TR(model)} into eq.~\ref{N(t)} we find 
\begin{equation}\label{3.20}
N(t) = \left(\int_0^{\infty} S(\lambda') \epsilon(\lambda') R(\lambda')
\exp\left(-nx\Big[ \sigma_0(\lambda') + \sigma_{hkl}(\lambda')(1-\mathcal{H}(\lambda' - 2d_{hkl}))\Big] \right)d\lambda'\right)\Delta t.
\end{equation}
We assume that $S(\lambda)$ and $\epsilon(\lambda)$ are `flat' (approximately constant) functions within the narrow integration window defined by $R(\lambda)$~\cite{Santisteban:2001}. Hence, we can take them outside the integral and eq.~\ref{3.20} becomes 
\begin{equation}
N(t) \approx \left(S(\lambda') \epsilon(\lambda')\exp(-nx\sigma_0(\lambda')) \int_0^{\infty} \exp\Big(-nx\sigma_{hkl}(\lambda')(1-\mathcal{H}(\lambda' - 2d_{hkl}))\Big) R(\lambda') d\lambda' \right) \Delta t.
\end{equation}

To proceed, we begin by normalising $N(t)$ by the number of neutrons counted with the sample removed $N_0(t)$ (\emph{i.e.} the number of emitted neutrons) such that 
\begin{equation}
M(t) = \frac{N(t)}{N_0(t)} \approx \frac{\Big(S(\lambda') \epsilon(\lambda')\exp(-nx\sigma_0(\lambda')) \int_0^{\infty} \exp\Big(-nx\sigma_{hkl}(\lambda')(1-\mathcal{H}(\lambda' - 2d_{hkl}))\Big) R(\lambda') d\lambda'  \Big)\Delta t}{\Big(S(\lambda') \epsilon(\lambda')\int_0^{\infty} R(\lambda') d\lambda' \Big) \Delta t}
\end{equation}
which simplifies to give 
\begin{equation}
M(t) \approx \exp(-nx\sigma_0(\lambda')) \int_0^{2d_{hkl}}  \exp(-nx\sigma_{hkl}(\lambda')) R(\lambda') d\lambda', 
\end{equation}
under the assumption that $R(\lambda')$ is normalised. We now note that 
$$
\int_0^{2d_{hkl}} \exp(-nx\sigma_{hkl}(\lambda)) R(\lambda) d\lambda = \int_0^{\infty}  \exp(-nx\sigma_{hkl}(\lambda)) R(\lambda) d\lambda - \int_{2d_{hkl}}^{\infty}  \exp(-nx\sigma_{hkl}(\lambda)) R(\lambda) d\lambda
$$
and following~\cite{Santisteban:2001} we assume that  $\exp(-nx\sigma_{hkl}(\lambda'))$ is a constant function in the narrow region defined by $R(\lambda')$; hence we yield
\begin{equation}\label{finalbeforeint}
M(t) \approx \exp(-nx\sigma_0(\lambda')) \left( 1- \exp(-nx\sigma_{hkl}(\lambda')) \right)\int_{2d_{hkl}}^{\infty}   R(\lambda') d\lambda'.
\end{equation}

Finally, we substitute the function $R(\lambda)$ as defined in eq.~\ref{Kropfff_Almost_done} into eq.~\ref{finalbeforeint} and by making use of integration by parts, we converge to:
\begin{equation}
M(t) \approx \exp(-nx\sigma_0(\lambda')) \left( 1- \exp(-nx\sigma_{hkl}(\lambda')) \right)B(\lambda)
\end{equation}
where $B(\lambda)$ is given by
\begin{equation}\label{B(lambda)}
B\left(\lambda \right)=
\frac{1}{2}\left[\operatorname{erfc}\left(-\frac{\lambda - \lambda_{hkl}}{\sqrt{2} \sigma}\right)-\exp \left(-\frac{\lambda - \lambda_{hkl}}{\alpha}+\frac{\sigma^{2}}{2 \alpha^{2}}\right) \operatorname{erfc}\left(-\frac{\lambda - \lambda_{hkl}}{\sqrt{2}\sigma}+\frac{\sigma}{\alpha}\right)\right].
\end{equation}
\noindent
In eq.~\ref{B(lambda)}, $\lambda_{hkl}$ represent the wavelength at the Bragg edge and all other terms have the same meaning as defined in eq.~\ref{Kropfff_Almost_done}. For the purposes of the fitting, a further simplifying assumption is made to the parametric fitting function; we approximate the arguments of the two exponential functions in eq.~\ref{finalbeforeint} as linear function of wavelength such that the final function becomes
\begin{equation}\label{FinalB}
M(t) \approx \exp(-(a_0 + b_0 \lambda))( 1- \exp(-(a_1 + b_1\lambda))) B(\lambda).
\end{equation}
Eq.~\ref{FinalB} provides a parametric model which we use in the main paper to perform a non-linear fit of the neutron spectra in the neighbourhood of a Bragg edge.

\printbibliography[heading=subbibliography]

\end{refsection}

%% file: mybib.bib
@PREAMBLE{
 "\providecommand{\noopsort}[1]{}" 
 # "\providecommand{\singleletter}[1]{#1}%" 
}

@article{abbey2009feasibility,
  title={Feasibility study of neutron strain tomography},
  author={Abbey, Brian and Zhang, Shu Yan and Vorster, Wim JJ and Korsunsky, Alexander M},
  journal={Procedia Engineering},
  volume={1},
  number={1},
  pages={185--188},
  year={2009},
  publisher={Elsevier},
  addendum={{DOI}: \href{https://doi.org/10.1016/j.proeng.2009.06.043}{10.1016/j.proeng.2009.06.043}}
}

@article{abbey2012neutron,
  title={Neutron strain tomography using {B}ragg-edge transmission},
  author={Abbey, Brian and Zhang, Shu Yan and Xie, Mengyin and Song, Xu and Korsunsky, Alexander M},
  journal={International journal of materials research},
  volume={103},
  number={2},
  pages={234--241},
  year={2012},
  publisher={De Gruyter},
  addendum={{DOI}: \href{https://doi.org/10.3139/146.110674}{10.3139/146.110674}}
}

@article{abbey2012reconstruction,
  title={Reconstruction of axisymmetric strain distributions via neutron strain tomography},
  author={Abbey, Brian and Zhang, Shu Yan and Vorster, Wim and Korsunsky, Alexander M},
  journal={Nuclear Instruments and Methods in Physics Research Section B: Beam Interactions with Materials and Atoms},
  volume={270},
  pages={28--35},
  year={2012},
  publisher={Elsevier},
  addendum={{DOI}: \href{https://doi.org/10.1016/j.nimb.2011.09.012}{10.1016/j.nimb.2011.09.012}}
}

@article{tremsin2011high,
  title={High resolution {B}ragg edge transmission spectroscopy at pulsed neutron sources: proof of principle experiments with a neutron counting MCP detector},
  author={Tremsin, AS and McPhate, JB and Kockelmann, W and Vallerga, JV and Siegmund, OHW and Feller, WB},
  journal={Nuclear Instruments and Methods in Physics Research Section A: Accelerators, Spectrometers, Detectors and Associated Equipment},
  volume={633},
  pages={S235--S238},
  year={2011},
  publisher={Elsevier},
  addendum={{DOI}: \href{https://doi.org/10.1016/j.nima.2010.06.176}{10.1016/j.nima.2010.06.176}}
}

@article{tremsin2012high,
  title={High Resolution Photon Counting With MCP-Timepix Quad Parallel Readout Operating at >1 KHz Frame Rates},
  author={Tremsin, Anton S and Vallerga, John V and McPhate, Jason B and Siegmund, Oswald HW and Raffanti, Rick},
  journal={IEEE transactions on nuclear science},
  volume={60},
  number={2},
  pages={578--585},
  year={2012},
  publisher={IEEE},
  addendum={{DOI}: \href{https://doi.org/10.1109/TNS.2012.2223714}{10.1109/TNS.2012.2223714}}
}

@article{lionheart2015diffraction,
  title={Diffraction tomography of strain},
  author={Lionheart, William RB and Withers, Philip J},
  journal={Inverse Problems},
  volume={31},
  number={4},
  pages={045005},
  year={2015},
  publisher={IOP Publishing},
  addendum={{DOI}: \href{https://doi.org/10.1088/0266-5611/31/4/045005}{10.1088/0266-5611/31/4/045005}}
}

@inproceedings{gregg2017bragg,
  title={Bragg-edge neutron strain tomography: {A} review and path forward to general tomographic reconstruction},
  author={Gregg, AWT and Hendriks, JN and Wensrich, CM and others},
  booktitle={9th Australasian Congress on Applied Mechanics (ACAM9)},
  pages={274--282},
  year={2017},
  organization={Engineers Australia Sydney},
  url = {https://search.informit.org/doi/10.3316/informit.392701260273286}
}

@article{lionheart2018histogram,
  title={Histogram tomography},
  author={Lionheart, William RB},
  journal={Mathematics in Engineering},
  pages={55-74},
  volume={2},
  issue={1},
  year={2020},
  addendum={{DOI}: \href{https://doi.org/10.3934/mine.2020004}{10.3934/mine.2020004}}
}

@article{daymond2002analysis,
  title={Analysis of neutron diffraction strain measurement data from a round robin sample},
  author={Daymond, MR and Johnson, MW and Sivia, DS},
  journal={The Journal of Strain Analysis for Engineering Design},
  volume={37},
  number={1},
  pages={73--85},
  year={2002},
  publisher={SAGE Publications Sage UK: London, England},
  addendum={{DOI}: \href{https://doi.org/10.1243/0309324021514844}{10.1243/0309324021514844}}
}

@article{lionheart2019tof,
	title={Can the second moment of the {B}ragg edge be resolved for neutron strain measurement? },
	author={Lionheart, William R. B. and Burca, Genoveva and Korsunsky, Alexander and Turner, Martin and J{\o}rgensen, Jakob Sauer and Schmidt, S{\o}ren and Kelleher, Joe and Yan, Kun and Withers, Philip J. and Ametova, Evelina},
	journal={STFC ISIS Neutron and MuonSource},
	year={2019},
	addendum = {{DOI}: \href{https://doi.org/10.5286/ISIS.E.100529645}{10.5286/ISIS.E.100529645}}
}

@article{suortti1979voigt,
  title={Voigt function fit of {X}-ray and neutron powder diffraction profiles},
  author={Suortti, P and Ahtee, M and Unonius, L},
  journal={Journal of Applied Crystallography},
  volume={12},
  number={4},
  pages={365--369},
  year={1979},
  publisher={International Union of Crystallography},
  addendum={{DOI}: \href{https://doi.org/10.1107/S002188987901270X}{10.1107/S002188987901270X}}
}

@article{kropff1982bragg,
  title={The {B}ragg lineshapes in time-of-flight neutron powder spectroscopy},
  author={Kropff, F and Granada, JR and Mayer, RE},
  journal={Nuclear Instruments and Methods in Physics Research},
  volume={198},
  number={2-3},
  pages={515--521},
  year={1982},
  publisher={Elsevier},
  addendum={{DOI}: \href{https://doi.org/10.1016/0167-5087(82)90293-9}{10.1016/0167-5087(82)90293-9}}
}

@article{santisteban2001time,
  title={Time-of-flight neutron transmission diffraction},
  author={Santisteban, JR and Edwards, L and Steuwer, A and Withers, PJ},
  journal={Journal of Applied Crystallography},
  volume={34},
  number={3},
  pages={289--297},
  year={2001},
  publisher={International Union of Crystallography},
  addendum={{DOI}: \href{https://doi.org/10.1107/S0021889801003260}{10.1107/S0021889801003260}}
}

@article{liptak2019developments,
  title={Developments towards {B}ragg edge imaging on the {IMAT} beamline at the {ISIS} pulsed neutron and muon source: {BEA}n software},
  author={Liptak, Alexander and Burca, Genoveva and Kelleher, Joe and Ovtchinnikov, Evgueni and Maresca, Jacob and Horner, Aled},
  journal={Journal of Physics Communications},
  volume={3},
  number={11},
  pages={113002},
  year={2019},
  publisher={IOP Publishing},
  addendum={{DOI}: \href{https://doi.org/10.1088/2399-6528/ab5575}{10.1088/2399-6528/ab5575}}
}

@phdthesis{boin2011developments,
  title={Developments towards the tomographic imaging of local crystallographic structures},
  author={Boin, M},
  school={The Open University, Milton Keynes},
  year={2011}
}

@article{kockelmann2018time,
  title={Time-of-flight neutron imaging on {IMAT@} {ISIS}: a new user facility for materials science},
  author={Kockelmann, Winfried and Minniti, Triestino and Pooley, Daniel E and Burca, Genoveva and Ramadhan, Ranggi and Akeroyd, Freddie A and Howells, Gareth D and Moreton-Smith, Chris and Keymer, David P and Kelleher, Joe and others},
  journal={Journal of Imaging},
  volume={4},
  number={3},
  pages={47},
  year={2018},
  publisher={Multidisciplinary Digital Publishing Institute},
  addendum={{DOI}: \href{https://doi.org/10.3390/jimaging4030047}{10.3390/jimaging4030047}}
}

@article{burca2013modelling,
  title={Modelling of an imaging beamline at the {ISIS} pulsed neutron source},
  author={Burca, G and Kockelmann, W and James, JA and Fitzpatrick, Michael E},
  journal={Journal of Instrumentation},
  volume={8},
  number={10},
  pages={P10001},
  year={2013},
  publisher={IOP Publishing},
  addendum={{DOI}: \href{https://doi.org/10.1088/1748-0221/8/10/P10001}{10.1088/1748-0221/8/10/P10001}}
}

@article{tremsin2014optimization,
  title={Optimization of {T}imepix count rate capabilities for the applications with a periodic input signal},
  author={Tremsin, AS and Vallerga, JV and McPhate, JB and Siegmund, OHW},
  journal={Journal of Instrumentation},
  volume={9},
  number={05},
  pages={C05026},
  year={2014},
  publisher={IOP Publishing},
  addendum={{DOI}: \href{https://doi.org/10.1088/1748-0221/9/05/C05026}{10.1088/1748-0221/9/05/C05026}}
}

@article{santisteban2002strain,
  title={Strain imaging by {B}ragg edge neutron transmission},
  author={Santisteban, Javier R and Edwards, Lyndon and Fitzpatrick, Mike E and Steuwer, Axel and Withers, Philip J and Daymond, Mark R and Johnson, Michael W and Rhodes, Nigel and Schooneveld, Erik M},
  journal={Nuclear Instruments and Methods in Physics Research Section A: Accelerators, Spectrometers, Detectors and Associated Equipment},
  volume={481},
  number={1-3},
  pages={765--768},
  year={2002},
  publisher={Elsevier},
  addendum={{DOI}: \href{https://doi.org/10.1016/S0168-9002(01)01256-6}{10.1016/S0168-9002(01)01256-6}}
}

@article{woracek2011neutron,
  title={Neutron {B}ragg-edge-imaging for strain mapping under in situ tensile loading},
  author={Woracek, R and Penumadu, D and Kardjilov, N and Hilger, A and Strobl, M and Wimpory, RC and Manke, I and Banhart, J},
  journal={Journal of Applied Physics},
  volume={109},
  number={9},
  pages={093506},
  year={2011},
  publisher={American Institute of Physics},
  addendum={{DOI}: \href{https://doi.org/10.1063/1.3582138}{10.1063/1.3582138}}
}

@article{tremsin2012strain,
  title={High-resolution strain mapping through time-of-flight neutron transmission diffraction with a microchannel plate neutron counting detector},
  author={Tremsin, AS and McPhate, JB and Steuwer, A and Kockelmann, W and M Paradowska, A and Kelleher, JF and Vallerga, JV and Siegmund, OHW and Feller, WB},
  journal={Strain},
  volume={48},
  number={4},
  pages={296--305},
  year={2012},
  publisher={Wiley Online Library},
  addendum={{DOI}: \href{https://doi.org/10.1111/j.1475-1305.2011.00823.x}{10.1111/j.1475-1305.2011.00823.x}}
}

@inproceedings{tremsin2014strain,
  title={High-resolution strain mapping through time-of-flight neutron transmission diffraction},
  author={Tremsin, Anton S and McPhate, Jason B and Vallerga, John V and Siegmund, Oswald HW and Kockelmann, Winfried and Paradowska, Anna and Zhang, Shu Yan and Kelleher, Joe and Steuwer, Axel and Feller, W Bruce},
  booktitle={Materials Science Forum},
  volume={772},
  pages={9--13},
  year={2014},
  organization={Trans Tech Publ},
  addendum={{DOI}: \href{https://doi.org/10.4028/www.scientific.net/MSF.772.9}{10.4028/www.scientific.net/MSF.772.9}}
}

@article{hendriks2020bayesian,
  title={Bayesian non-parametric {B}ragg-edge fitting for neutron transmission strain imaging},
  author={Hendriks, Johannes and O’Dell, Nicholas and Wills, Adrian and Tremsin, Anton and Wensrich, Christopher and Shinohara, Takenao},
  journal={The Journal of Strain Analysis for Engineering Design},
  pages={0309324720959237},
  year={2020},
  publisher={SAGE Publications Sage UK: London, England},
  addendum={{DOI}: \href{https://doi.org/10.1177/0309324720959237}{10.1177/0309324720959237}}
}

@book{jaynes2003probability,
  title={Probability theory: {T}he logic of science},
  author={Jaynes, Edwin T},
  year={2003},
  publisher={Cambridge university press}
}

@article{betancourt2017conceptual,
  title={A conceptual introduction to {H}amiltonian {M}onte Carlo},
  author={Betancourt, Michael},
  journal={arXiv preprint arXiv:1701.02434},
  year={2017},
  addendum={{arXiv}: \href{https://arxiv.org/pdf/1701.02434.pdf}{arXiv:1701.02434}}
}

@article{d1971omnibus,
  title={An omnibus test of normality for moderate and large size samples},
  author={d'Agostino, Ralph B},
  journal={Biometrika},
  volume={58},
  number={2},
  pages={341--348},
  year={1971},
  publisher={Oxford University Press},
  addendum={{DOI}: \href{https://doi.org/10.1093/biomet/58.2.341}{10.1093/biomet/58.2.341}}
}

@article{webster2001polycrystalline,
  title={Polycrystalline materials. {D}eterminations of residual stresses by neutron diffraction},
  author={Webster, GA and Wimpory, RW},
  journal={ISO/TTA3 Technology Trends Assessment, Geneva},
  volume={20},
  year={2001}
}

@article{withers2001residual,
  title={Residual stress. {P}art 1--measurement techniques},
  author={Withers, Philip J and Bhadeshia, HKDH},
  journal={Materials science and Technology},
  volume={17},
  number={4},
  pages={355--365},
  year={2001},
  publisher={Taylor \& Francis},
  addendum={{DOI}: \href{https://doi.org/10.1179/026708301101509980}{10.1179/026708301101509980}}
}

@article{ramadhan2019characterization,
  title={Characterization and application of {B}ragg-edge transmission imaging for strain measurement and crystallographic analysis on the {IMAT} beamline},
  author={Ramadhan, Ranggi S and Kockelmann, Winfried and Minniti, Triestino and Chen, Bo and Parfitt, David and Fitzpatrick, Michael E and Tremsin, Anton S},
  journal={Journal of Applied Crystallography},
  volume={52},
  number={2},
  pages={351--368},
  year={2019},
  publisher={International Union of Crystallography},
  addendum={{DOI}: \href{https://doi.org/10.1107/S1600576719001730}{10.1107/S1600576719001730}}
}

@article{Santisteban:2001,
author = {Santisteban, J. R. and Edwards, L. and Steuwer, A. and Withers, P. J.},
title = "{Time-of-flight neutron transmission diffraction}",
journal = "Journal of Applied Crystallography",
year = "2001",
volume = "34",
number = "3",
pages = "289--297",
month = "Jun",
addendum = {{DOI}: \href{https://doi.org/10.1107/S0021889801003260}{10.1107/S0021889801003260}}
}

@article{Kropff1982,
  title={The bragg lineshapes in time-of-flight neutron powder spectroscopy},
  author={Kropff, F and Granada, JR and Mayer, RE},
  journal={Nuclear Instruments and Methods in Physics Research},
  volume={198},
  number={2-3},
  pages={515--521},
  year={1982},
  publisher={Elsevier},
  addendum = {{DOI}: \href{https://doi.org/10.1016/0167-5087(82)90293-9}{10.1016/0167-5087(82)90293-9}}
}

@article{Suortti1979,
  title={Voigt function fit of X-ray and neutron powder diffraction profiles},
  author={Suortti, P and Ahtee, M and Unonius, L},
  journal={Journal of Applied Crystallography},
  volume={12},
  number={4},
  pages={365--369},
  year={1979},
  publisher={International Union of Crystallography},
  addendum = {{DOI}: \href{https://doi.org/10.1107/S002188987901270X}{10.1107/S002188987901270X}},
}

@article{Vogel2000,
author = {Vogel, Sven},
journal = {PhD thesis},
title = {{A Rietveld-Approach for the Analysis of Neutron Time-Of-Flight Transmission Data}},
year = {2000},
url={https://macau.uni-kiel.de/servlets/MCRFileNodeServlet/dissertation_derivate_00000330/d330.pdf}
}

@book{webster2000neutron,
	title={Neutron diffraction measurements of residual stress in a shrink-fit ring and plug},
	author={Webster, George A},
	year={2000},
	publisher={National Physical Laboratory}
}

@article{boin2011validation,
	title={Validation of Bragg edge experiments by Monte Carlo simulations for quantitative texture analysis},
	author={Boin, M and Hilger, A and Kardjilov, N and Zhang, SY and Oliver, EC and James, JA and Randau, C and Wimpory, RC},
	journal={Journal of Applied Crystallography},
	volume={44},
	number={5},
	pages={1040--1046},
	year={2011},
	publisher={International Union of Crystallography}
}

@phdthesis{Vogel_2000, 
	title={A Rietveld-Approach for the Analysis of Neutron Time-of-Flight Transmission Data}, 
	url={https://macau.uni-kiel.de/receive/diss_mods_00000330}, 
	author={Vogel, Sven}, 
	year={2000} }

@book{fitzpatrick2003analysis,
	title={Analysis of residual stress by diffraction using neutron and synchrotron radiation},
	author={Fitzpatrick, Michael E and Lodini, Alain},
	year={2003},
	publisher={CRC Press}
}

@article{steuwer2001sin2,
	title={The sin2 $\psi$-method in pulsed neutron transmission},
	author={Steuwer, A and Withers, PJ and Santisteban, JR and Edwards, L and Fitzpatrick, ME and Daymond, MR and Johnson, MW},
	journal={Journal of Neutron Research},
	volume={9},
	number={2-4},
	pages={289--294},
	year={2001},
	publisher={Taylor \& Francis}
}

@article{losko2021new,
	title={New perspectives for neutron imaging through advanced event-mode data acquisition},
	author={Losko, AS and Han, Y and Schillinger, B and Tartaglione, A and Morgano, M and Strobl, M and Long, J and Tremsin, AS and Schulz, M},
	journal={Scientific reports},
	volume={11},
	number={1},
	pages={1--11},
	year={2021},
	publisher={Nature Publishing Group}
}
